\documentclass[aps,showpacs]{revtex4}
\usepackage{psfig}

\begin{document}
%preprint number:
%
\title{
\[ \vspace{-2cm} \]
\noindent\hfill\hbox to 1.5in{\rm  } \vskip 1pt \noindent\hfill\hbox
to 1.5in{\rm SLAC-PUB-11260 \hfill  } \vskip 1pt
\noindent\hfill\hbox to 1.5in{\rm September 29, 2005 \hfill}\vskip 10pt
Adaptive Perturbation Theory I: Quantum Mechanics\footnote{This work
was supported by the U.~S.~DOE, Contract No.~DE-AC03-76SF00515.}}
\author{Marvin Weinstein}
\address{Stanford Linear Accelerator Center, Stanford University,
  Stanford, California 94309}
\date{September 29, 2005}
\begin{abstract}
Adaptive perturbation is a new method for perturbatively computing
the eigenvalues and eigenstates of quantum mechanical Hamiltonians
that heretofore were not believed to be obtainable by such methods.
The novel feature of adaptive perturbation theory is that it
decomposes a given Hamiltonian, $H$, into an unperturbed part and a
perturbation in a way which extracts the leading non-perturbative
behavior of the problem exactly.  This paper introduces the method
in the context of the pure anharmonic oscillator and then goes on to
apply it to the case of tunneling between both symmetric and
asymmetric minima.  It concludes with an introduction to the
extension of these methods to the discussion of a quantum field
theory.  A more complete discussion of this issue will be given in
the second paper in this series.  This paper will show how to use
the method of adaptive perturbation theory to non-perturbatively
extract the structure of mass, wavefunction and coupling constant
renormalization.
\end{abstract}
% slac pub 10244
\maketitle

\newcommand{\ba}{\begin{eqnarray}}
\newcommand{\ea}{\end{eqnarray}}
\newcommand{\x}{\mbox{$\vec{x}$}}
\newcommand{\be}{\begin{equation}}
\newcommand{\ee}{\end{equation}}
\def\ket#1{\vert #1 \rangle}
\def\bra#1{\langle #1 \vert}
\def\vev#1{\left< #1 \right>}
\newcommand{\Adag}{{A^{\dag}_\gamma}}
\newcommand{\A}{{A_\gamma}}
\newcommand{\N}{{N_\gamma}}
\newcommand{\n}{{n_\gamma}}
\newcommand{\ov}{\langle \chi(-c,\gamma) \vert \chi(c,\gamma) \rangle}
%\narrowtext

\section{Introduction}

Perturbation theory, one of the most useful tools in the physicist's
toolbox, is the method most commonly used to analyze problems in
quantum mechanics and field theory.  It is so useful that it is
applied even when the perturbation parameter isn't small enough for
the expansion to converge.  One then typically tries to extract
meaningful results by applying some summation technique to the
resulting series, often with mixed success.  What is really needed
is a method that extracts the leading non-perturbative behavior of
the quantity to be computed, and then corrects this leading
approximation in a systematic manner. This paper introduces a simple
technique that works in just this way.  I call this method {\it
adaptive perturbation theory\/}\cite{Feranchuketal}.

I will limit the bulk of this paper to discussing how adaptive
perturbation theory works on problems in quantum mechanics.
In the second paper in this series I will show how to apply an
appropriately generalized version of adaptive perturbation theory to
a quantum field theory.  In particular, I will then show how to
extract, in a non-perturbative context, the general structure of
mass, wavefunction and coupling constant renormalization.

\section{Example 1: The Pure Anharmonic Oscillator}

To introduce the ideas behind {\it adaptive perturbation theory}, I
begin by studying the pure anharmonic oscillator since, naively, this
problem seems to be impossible to treat perturbatively because it
is devoid of a small expansion parameter.  The pure anharmonic
oscillator is defined by a Hamiltonian of the form
\be
    H = \frac{1}{2}\, p^2 + \frac{1}{6}\,\lambda\, x^4 ,
\label{anharmone}
\ee
where the operators $x$ and $p$ satisfy the usual canonical
commutation relation $[x,p] = i$.

Since $H$ contains only the terms $p^2$ and $\lambda\,x^4$, there
does not seem to be a way to perturbatively compute the eigenstates
and eigenvalues of $H$ based upon a division of $H$ into an
unperturbed Hamiltonian $H_0$, whose eigenstates are known, and a
perturbation $V$.  Nevertheless, I will now show that this can be
done. The trick is to define a decomposition of the Hamiltonian in
Eq.~\ref{anharmone} into an unperturbed (i.e., solvable) part and a
perturbation, in a way that doesn't depend upon $\lambda$.  The choice
of decomposition is what is {\it adapted\/} to the problem
at hand.  Thus, the decomposition, and therefore the
perturbation theory for each eigenstate, will be different.

To begin the process of decomposing $H$, I introduce a one-parameter
family of annihilation and creation operators, $\Adag$ and $\A$, as follows:
\ba
    x &=& {1 \over \sqrt{2 \gamma}}\left( \Adag + \A \right) , \nonumber\\
    p &=& i\,\sqrt{\gamma \over 2} \left( \Adag - \A \right).
\ea
Note that given the commutation relations of $x$ and $p$, the commutation
relation $[ \A, \Adag ] = 1 $ holds independent of $\gamma$.
In terms of these operators $H$ can be rewritten as
\ba
    H &=& \frac{\gamma}{4} \left( -\Adag^2 -\A^2 + 2\,\Adag \A + 1 \right) \nonumber\\
    &+& \frac{\lambda}{4\gamma^2} \left( \Adag^2 \A^2 +  2 \Adag \A + 1
    + \frac{1}{6} (\Adag^4 + \A^4 )
    + \frac{2}{3} (\Adag^3 \A + \Adag \A^3 ) + ( \Adag^2 + \A^2 ) \right) \nonumber\\
    &=& \left( \frac{\gamma}{4} + \frac{\lambda}{4\gamma^2} \right) ( 2\,\N + 1)
    + \frac{\lambda}{4\gamma^2} \N\,(\N-1)
    + \left( \frac{\gamma}{4} - \frac{\lambda}{2\gamma^2} \right) (\Adag^2 + \A^2) \nonumber\\
    &+& \frac{\lambda}{4\gamma^2} \left( \frac{1}{6}(\Adag^4 + \A^4)
    + \frac{2}{3} (\Adag^3 \A + \Adag \A^3) \right)
\ea
where I have defined the $\gamma$-dependent number operator to be
\be
    \N = \Adag \A .
\ee
At this point I define a $\gamma$-dependent decomposition of $H$
into an unperturbed part and a perturbation as follows:
\ba
    H &=& H_0(\gamma) + V(\gamma), \label{pertdecompone} \nonumber\\
    H_0(\gamma) &=& \left( \frac{\gamma}{4} + \frac{\lambda}{4\gamma^2} \right) ( 2\,\N + 1)
    + \frac{\lambda}{4\gamma^2} \N\,(\N-1) ,  \nonumber\\
    V(\gamma) &=& \left( \frac{\gamma}{4} - \frac{\lambda}{2\gamma^2} \right) (\Adag^2 + \A^2)
      \nonumber\\
    &+& \frac{\lambda}{4\gamma^2} \left( \frac{1}{6}(\Adag^4 + \A^4)
    + \frac{2}{3} (\Adag^3 \A + \Adag \A^3) \right) .
\label{pertdecomplast}
\ea
In other words, the unperturbed Hamiltonian is defined to be that
part of $H$ which is diagonal in the number operator $\N$ and the
perturbation is defined to be everything else.

If we think of the Hamiltonian as a large matrix in the basis of
Fock-space states generated by applying $\Adag$ to the state
$\ket{0_\gamma}$, then $H_0(\gamma)$ is the diagonal part of the
matrix and the perturbation is the matrix obtained by setting all
terms on the diagonal to zero.

Although it isn't conventional to include the term
$\frac{\lambda}{4\gamma^2} \N\,(\N-1)$ in the definition of the
unperturbed Hamiltonian, $H_0(\gamma)$, it makes sense to do so for
two reasons. First, this part of the Hamiltonian is diagonal in $\N$
and thus it is just as easy to deal with as the part proportional to
the first power of $\N$.  Second, including it in the unperturbed
Hamiltonian puts a term proportional to $\N^2$ in the energy
denominators which appear in the perturbation expansion based upon
this decomposition.  This suppresses the $n!$ growth of the usual
perturbation expansion. The only apparent drawback to putting a term
proportional to $\N^2$ into $H_0(\gamma)$ is that it makes the
energies of the unperturbed levels grow like $n^2$, whereas the
correct answer is $n^{4/3}$. While this is not a disaster, it seems
to say that getting the correct answer from a straightforward
perturbative approach will be difficult.  This observation leads us
to add the {\it adaptive\/} element to {\it adaptive perturbation
theory\/}.

The essential point is that I have not as yet committed to
a particular value of the parameter $\gamma$ in
Eqs.~\ref{pertdecompone}-\ref{pertdecomplast}. The trick consists of
adapting the choice of $\gamma$ to the computation at hand.  Thus,
to compute the energy of the $N^{th}$ eigenstate I will choose a
value, $\gamma(N)$, which varies as a function of $N$. The
appropriate value of $\gamma(N)$ is fixed by a simple variational
calculation.

The variational calculation, which serves as the starting
point for any adaptive perturbation theory computation, begins by
defining a $\gamma$-dependent family of Fock-states, $\ket{\N}$.
First we define the $\gamma$-dependent vacuum state,
$\ket{0_\gamma}$, by the condition
\be
    \A \ket{0_\gamma} = 0 .
\ee
Next, we define the $n$-particle state built on this vacuum state such that
\be
    N_\gamma \ket{\n} = n \ket{\n},
\ee
by
\be
    \ket{\n} = \frac{1}{\sqrt{n!}} \,\Adag^n \ket{0_\gamma} .
\ee

In terms of functions of $x$, the state $\ket{0_\gamma}$ is just a
gaussian whose width is determined by $\gamma$.  Similarly, the
state $\ket{\n}$ is the appropriate $n^{th}$ order Hermite
polynomial obtained by applying the $\gamma$-dependent differential
operator \be
    \Adag = \sqrt{\gamma \over 2}\,x - \frac{1}{\sqrt{2\gamma}}\,\frac{d}{dx}
\ee to this gaussian. (It is worth noting that the specific
functional form of the state $\ket{\n}$ depends upon $\gamma$, but
the number of times this function takes the value zero does not.)

The value of $\gamma$ used to define the adaptive perturbation
theory for the $n^{th}$ level of the anharmonic oscillator is determined by
requiring that it minimize the expectation value
\be
    E_n(\gamma) = \bra{\n} H \ket{\n} .
\label{engamma}
\ee
Eqs.~\ref{pertdecompone}-\ref{pertdecomplast} show that
this expectation value is equal to
\be
    E_n(\gamma) = \bra{\N} H_0(\gamma) \ket{n} = \left( \frac{\gamma}{4}
    + \frac{\lambda}{4\gamma^2} \right) ( 2\,n + 1)
    + \frac{\lambda}{4\gamma^2} n\,(n-1).
\ee
Minimizing $E_n(\gamma)$ with respect to $\gamma$ leads to the equation
\be
    \gamma = \lambda^{1/3}\left( \frac{2\,(n^2 + n +1)}{2n+1} \right)^{1/3} .
\ee
At this point all we have to do is substitute this value into Eq.~\ref{engamma}, to
see why we chose $\gamma$ to minimize $E_n(\gamma)$.  If we do this, we obtain
\be
    E_n(\gamma)_{\rm min} = \frac{3}{8}\,\lambda^{1/3} \left(2 n + 1 \right) ^{2/3}
    \left(2 n^2 + 2(n+1) \right)^{1/3}
\ee
which, for large $n$, behaves as $\lambda^{1/3}\,n^{4/3}$ , which is the correct
answer.

The fact that all energies scale as $\lambda^{1/3}$ is an exact
result and so the non-trivial part of variational computation is the
derivation of the dependence of the energy on $n$. To see why all energies
are proportional to $\lambda^{1/3}$, it suffices to make the following
canonical transformation
\be
    x \rightarrow \frac{x}{\lambda^{1/6}} \quad ; \quad p \rightarrow \lambda^{1/6} p.
\ee
In terms of these operators, the Hamiltonian of the pure anharmonic oscillator becomes
\be
    H = \lambda^{1/3} \left( \frac{1}{2} p^2 + \frac{1}{6} x^4 \right),
\ee
thus proving the claim.  A comparison of the variational computation and the
result of a second-order perturbation theory for $\lambda = 1$ and widely differing values of
$N$ is given in Table \ref{tableone}.  As advertised, the
adaptive perturbation theory for each level converges rapidly, independent of
$\lambda$ and $N$.

\begin{table}[b!]
\caption{A comparison of the zeroth order and second order perturbation results, for the
energy of the $N^{th}$ level of the pure anharmonic oscillator, to the exact answer for $\lambda = 1$
and widely varying values of $N$.}
\vskip 5pt
\label{anharmtable}
\begin{center}
\begin{tabular}{|c|c|c|c|c|c|c|} \hline
$\lambda$ &  $N$ & Variational & $2^{nd}$ Order Perturbation & Exact & Variational \% Err & Perturbative \% Error \\
\hline
1.0 & 0  & 0.375 & 0.3712 & 0.3676 & 0.02 & 0.0098 \\
\hline
1.0 & 1  & 1.334   & 1.3195 & 1.3173   & 0.01   & 0.0017 \\
\hline
1.0 & 10 & 17.257  & 17.508 &  17.4228  & -0.009 & 0.0049 \\
\hline
1.0 & 40 & 104.317 & 105.888 & 105.360  & -0.009 & 0.0050 \\
\hline
\end{tabular}
\end{center}
\label{tableone}
\end{table}

\subsection{Why does the expansion converge?}

At this point I should address the question of why adaptive
perturbation theory converges.

First, observe that once I have done the variational calculation, the
perturbation theory computation is done for states defined in terms
of a single fixed value of $\gamma$. Thus, all states which appear
in the perturbative expansion are orthogonal to one another and the
perturbation series is well-defined.  Second, for the values of
$\gamma$ associated with a given state $\ket{n_0}$, the unperturbed
energies for states having another occupation number, $n$,
all have a term $n^2/\gamma^2 \approx
n^2/n_0^{2/3}$. It is only for values of $n$ that are near $n_0$
that the energy denominators behave as $n^{4/3}$.  For values of $n >>
n_0$ the factor $\gamma$ does not suppress the $n^2$ behavior.
This is what suppresses the $n!$-growth of the perturbation
expansion, since now the factor of $\sqrt{(n+1)(n+2)(n+3)(n+4)}$,
which comes from applying $\Adag^4$ to a state $\ket{\N}$, is canceled
by the $1/n^2$ appearing in the energy denominator,
instead of the usual $1/n$.  I'll make no attempt to prove this
assertion, instead I refer the reader to a pair of very nice papers
by I.~G.~Halliday and P.~Suranyi\cite{Halliday:1979vn,
Halliday:1979xh} where, by a very different technique which shares
this same feature, they give a proof that the fact that the
unperturbed energies are quadratic in $N$ leads to a convergent
perturbation expansion.

Finally, I should comment that the fact that variationally
determined states $\ket{\n}$ are not orthogonal to one another
isn't a problem.
First, there is a simple formula that expresses the state
$\ket{0_\gamma}$ as a sum over all states $\ket{n_{\gamma'}}$ when
$\gamma'$ different from $\gamma$ (see
Appendix A).  Thus, if one wishes to do a non-perturbative diagonalization
of the full Hamiltonian truncated to many different variationally
determined states, one can either orthogonalize these states by hand, or
use the formula for the overlap of these states and solve the so-called
{\it relative eigenvalue problem\/}
\be
    H \ket{\psi} = \epsilon \, M \ket{\psi}.
\ee
Note, that if $\ket{n_{\gamma(n)}}$ stands for one of the variationally
constructed states, then the Hermitian matrix $M$ is given by
\be
    M_{n,m} = \langle M_{\gamma(n)}\vert M_{\gamma(m)} \rangle .
\ee
The second point is that since the perturbation expansion for any state
$\ket{n_{\gamma(n)}}$ converges, eventually the perturbatively corrected
states $\ket{n_{\gamma(n)}}$ become orthogonal to one another.  This follows
from the fact that eigenstates of a Hermitian Hamiltonian having distinct
eigenvalues are orthogonal to one another.

\subsection{What does this have to do with quasi-particles?}

An interesting corollary to the adaptive perturbation theory
technique is that it provides an explicit realization of the
quasi-particle picture underlying much of many-body theory.  What we
have shown is that no matter how large the underlying coupling, the
physics of the states near a given $n_0$-particle state can be
accurately described in terms of perturbatively coupled eigenstates
of an appropriately chosen harmonic oscillator.  Of course, as we
have seen, the appropriately chosen harmonic oscillator picture
changes as $n_0$ changes.  Furthermore, the perturbatively coupled
states $\ket{n_{\gamma(n_0)}}$, for $n \approx n_0$, correspond to states
containing an infinite number of particles, if we choose as a basis
those states that correspond to a significantly different value of
$n_0$.

\section{Example 2:  Adding a Mass Term}

An obvious modification of the pure anharmonic oscillator is to add
a positive mass term; i.e., let $H$ be
\be
    H = \frac{1}{2}\, p^2 + \frac{m^2}{2} x^2 + \frac{1}{6}\,\lambda\, x^4 .
\label{anharmtwo}
\ee
This problem is dealt with in exactly the same way.  First, introduce $\gamma$-dependent annihilation
and creation operators, to arrive at a slightly modified
formula for the expectation value in state $\ket{\N}$; {\it i.e.},
\be
    E_n(\gamma) =  \left( \frac{\gamma}{4}
    + \frac{m^2}{4\gamma} + \frac{\lambda}{4\gamma^2} \right) ( 2\,n + 1)
    + \frac{\lambda}{4\gamma^2} n\,(n-1).
\label{varen}
\ee
Once again, minimizing $E_n(\gamma)$ with respect to $\gamma$ leads to the equation,
\be
    \gamma^3 - m^2 \gamma - 2\lambda \left(\frac{n^2+n+1}{2 n + 1} \right) = 0 .
\ee
Clearly, for small $\lambda$ and small $n$, the solution to this equation is
$\gamma \approx m$.  In other words, $\gamma$ is what it would be for the usual
perturbative expansion.  Note, however, that no matter how small $\lambda $ is,
the term $\lambda n(n-1)$ eventually dominates the energy denominator
and shuts off the $N!$-growth of the perturbation expansion.  Also, for large enough
$n$, it is clear that Eq.~\ref{varen} is eventually dominated by the term proportional
to $\lambda$, and for those states the energy is proportional to
$\lambda^{1/3} n^{4/3}$.

As in the previous case, numerical studies show that independent of $\lambda$, $m$ and
$n$ the second order perturbation theory determined by the variational calculation
gives the eigenvalues of $H$ to an accuracy of better than one percent.

\section{Example 3:  The Double Well}

It should be clear from the previous discussion that adaptive perturbation theory
can be applied to any Hamiltonian of the form
\be
   H = \frac{p^2}{2} + \frac{m^2}{2}x^2 + \lambda\,x^p. \ee The
problem becomes more interesting when the mass term is chosen to be
negative and the potential energy develops two distinct minima.
(This is interesting in the case of a field theory because it is
associated with the sponataneous breaking of the discrete symmetry
$\phi \rightarrow -\phi$.)  I will now show how to apply adaptive
perturbation theory to this problem, since obtaining
accurate answers requires generalizing the procedure in a way that
will be very important in the next paper in this series.

Up to an irrelevant constant, the most general negative mass version
of the anharmonic oscillator can be written as:
\be
    H = \frac{1}{2}\, p^2 + \frac{1}{6}\,\lambda\, \left( x^2 - f^2 \right)^2 .
\label{doublewell}
\ee
Clearly, for a non-vanishing value of $f^2$, this potential has two minima
located at $x = \pm f$.  Thus, we would expect that the best
gaussian approximation to the ground state of this system can't be a
gaussian centered at the origin; classical intuition would imply that it
is a gaussian centered about another point, $x=c$.
In other words, if $\ket{0_\gamma}$ is a state centered
at $x=0$, it is better to adopt a trial state of the form
\be
    \ket{c,\gamma} = e^{-i c p}\,\ket{0_\gamma} .
\label{shiftedstate}
\ee
Since
\be
    e^{i c p}\, x\, e^{-i c p} = x + c ,
\ee
it follows that computing the expectation value of the Hamiltonian, Eq.~\ref{doublewell},
in the state specified in Eq.~\ref{shiftedstate}, is the same as computing the expectation
value of the Hamiltonian obtained by replacing the operator $x$ by $x+c$, in the state
$\ket{0_\gamma}$.  In the language of the previous sections, this is equivalent to introducing
the annihilation and creation operators $\Adag$ and $\A$ as follows:
\ba
    x &=& {1 \over \sqrt{2 \gamma}}\left( \Adag + \A \right) + c , \nonumber\\
    p &=& i\,\sqrt{\gamma \over 2} \left( \Adag - \A \right).
\label{shiftedops}
\ea
With this definition, the normal ordered form of the Hamiltonian, Eq.~\ref{doublewell}, becomes
\ba
    H &=& H(\gamma,c) = \left( \frac{\gamma}{4} + \frac{\lambda f^4}{6} - \frac{\lambda f^2 c^2}{3}
    + \frac{\lambda c^4}{6} + \frac{\lambda c^2}{2 \gamma} + \frac{\lambda}{8 \gamma^2}
    - \frac{\lambda f^2}{6 \gamma} \right)
    + \frac{\lambda c}{3} \sqrt{\frac{2}{\gamma}} \left( c^2 - f^2 + \frac{3}{2 \gamma}
    \right) \left( \Adag + \A \right) \nonumber\\
    & + & \left( \frac{\lambda c^2}{\gamma} + \frac{\lambda}{2 \gamma^2} + \frac{\gamma}{2}
    - \frac{\lambda f^2}{ 3 \gamma} \right) \N + \frac{\lambda}{4 \gamma^2} \N ( \N-1)
    + \left( -\frac{\lambda f^2}{ 6 \gamma}
    + \frac{\lambda}{4 \gamma^2} + \frac{\lambda c^2}{2 \gamma} - \frac{\gamma}{4} \right)
    \left( \Adag^2 + \A^2 \right)
    \nonumber \\
    & + & \frac{\lambda c}{ 3 \sqrt{2} \gamma^{3/2}} \left( \Adag^3 + \A^3 \right)
    + \frac{\lambda}{6 \gamma^2} \left( \Adag^3 \A + \Adag \A^3 \right)
    + \frac{\lambda}{24 \gamma^2} \left( \Adag^4 + \A^4 \right) .
\label{normalorderedH}
\ea
The expectation value of this Hamiltonian in the state $\ket{0_\gamma}$ is
\be
    {\cal E}(c,\gamma) =\left( \frac{\gamma}{4} + \frac{\lambda f^4}{6} - \frac{\lambda f^2 c^2}{3}
    + \frac{\lambda c^4}{6} + \frac{\lambda c^2}{2 \gamma} + \frac{\lambda}{8 \gamma^2}
    - \frac{\lambda f^2}{6 \gamma} \right) ,
\ee
which should be minimized with respect to both $\gamma$ and $c$ in order to define
the starting point of the adaptive perturbation theory computation.

\begin{figure}[ht]
\begin{center}
\leavevmode
\psfig{file=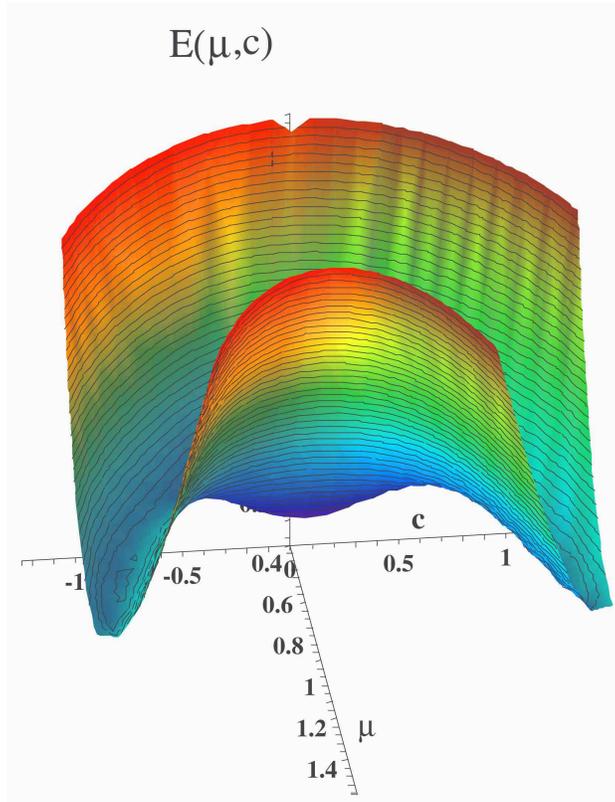,width=3.25in}
\end{center}
\caption{Effective potential as a function of $\gamma$ and $c$ showing
three local minima.}
\label{tricritthreed}
\end{figure}

\begin{figure}[ht]
\begin{center}
\leavevmode
\psfig{file=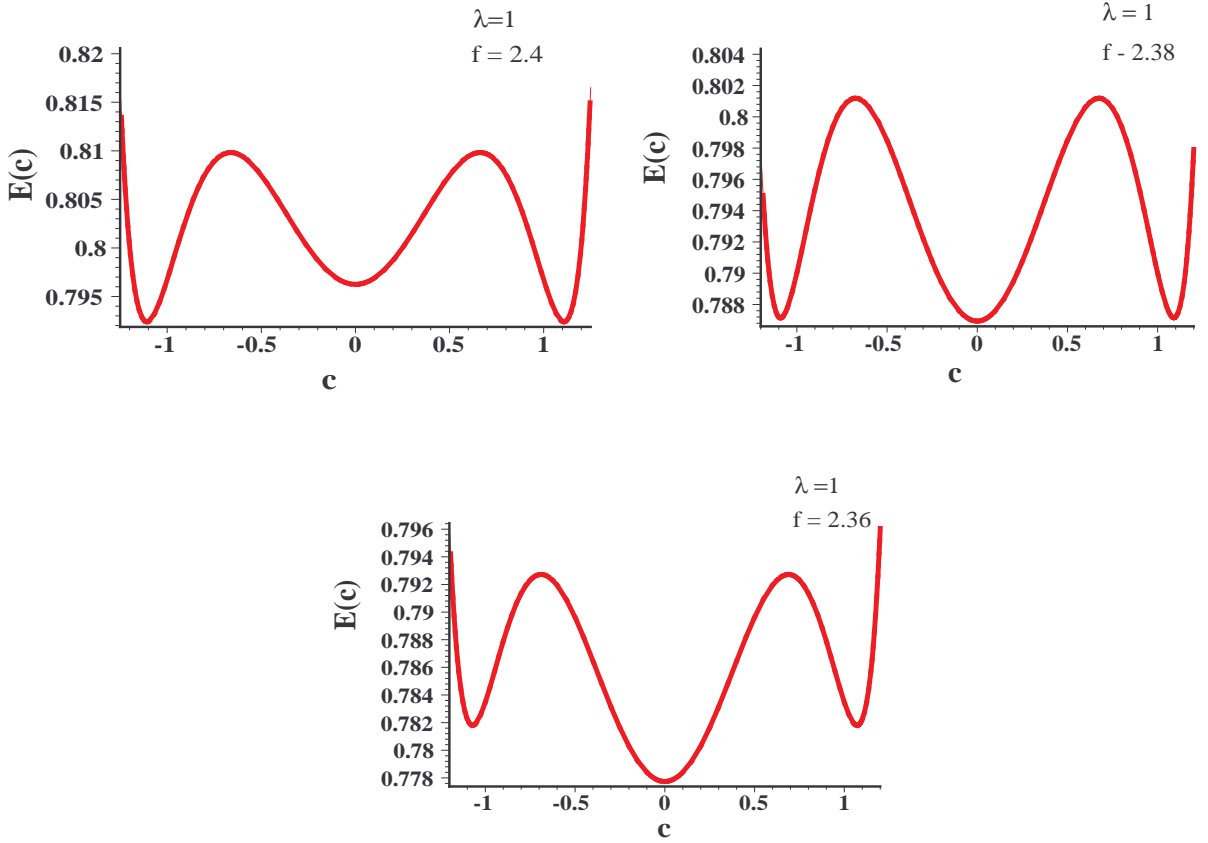,width=6.5in}
\end{center}
\caption{Effective potential as a function of $c$, showing two degenerate local minima
at $c=\pm c_{\rm min}$ and one minimum at $c=0$. The three plots are for $\lambda=1$ and different
values of $f$ and show how the minimum at $c=0$ changes from lying higher than the minima at
$c=\pm c_{\rm min}$ to being the global minimum.  If this were field theory this would be
characteristic of a first order phase transition.}
\label{tricritone}
\end{figure}

Fig.~\ref{tricritthreed} shows ${\cal E}(\gamma, c)$ for a particular choice of
$\lambda$ and $f^2$ and it exhibits what is a generic feature of any such plot: namely that
there are three minima.  To get a better feeling for what the three dimensional plot
is showing us, it is convenient to hold $c$ fixed and solve for the value of $\gamma$ that
minimizes ${\cal E}(\gamma, c)$, which I will denote by $\gamma(c)$, and then plot
${\cal E}(\gamma(c), c)$ for various values of $\lambda$ and $f^2$.
Three such plots are shown in
Fig.~\ref{tricritone}.  The first plot is for a value of $f$ which is large enough so that the
lowest energy is obtained for a gaussian shifted either to the right or left by the amount
$c = \pm c_{\rm min}$.  While there is a local minimum at $c=0$ it has a higher energy than the
shifted states.  Things change as one lowers the value of $f$.  Thus, the second plot shows
that for lower $f$ the shifted wavefunctions and the one centered at zero are essentially
degenerate in energy.  For a slightly smaller value of $f$, things reverse and the unshifted
wavefunction has a lower energy than the shifted ones.

While there would seem to be nothing wrong with the situation shown
in Fig.~\ref{tricritone}, it is problematic when one applies the same
sort of analysis to negative mass $\phi^4$ field theory in
$1+1$-dimensions.  In this case, if
one were to add a term like $J \phi$ to the Hamiltonian, this
result would imply the existence of a first order phase transition when $J$
reached some finite value.  At this point
the expectation value of $\phi$ in the ground-state would jump
discontinuously from zero to a non-zero value. It has been
rigorously shown that such a first order phase transition at a non-vanishing
value of $J$ cannot occur\cite{Simonetal,variational}.

\begin{figure}[b!]
\begin{center}
\leavevmode
\psfig{file=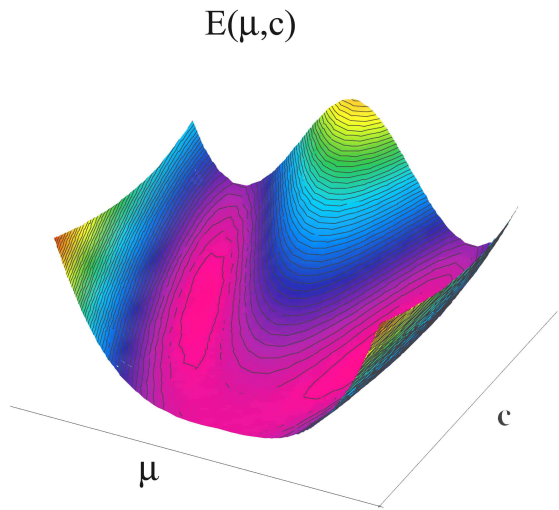,width=3in}
\end{center}
\caption{Effective potential as a function o f $c$, one global minimum.}
\label{shiftedgaussone}
\end{figure}

For this reason it is important to understand the origin of the
secondary minimum at $c=0$ and how to avoid it ever becoming lower
in energy than the flanking minima at $c = \pm c_{\rm min}$.
Fortunately the origin of the problem is easy to understand and
fixing it is not difficult. The key is that the Hamiltonian in
Eq.~\ref{normalorderedH} contains a term linear in $c$ and linear in
$\Adag + \A$.  Taking the effect of this term into account
removes - or at least minimizes - the importance of, the secondary
minimum at $c = 0$.  In order to allow this term to play a role
in the variational calculation, we have to adopt a slightly more general form
of the variational wavefunction; namely, we try a state of the form
\be
     \psi(\gamma,c,\alpha) =
     e^{ i c p} \left( \cos(\alpha)\, \ket{0_\gamma} + \sin(\alpha)\, \ket{1_\gamma} \right).
\ee
The easiest way to minimize the expectation value of the Hamiltonian, Eq.~\ref{normalorderedH},
in this state is to fix the values of $\gamma$ and $c$ and solve for the value
of $\alpha$ which minimizes this expression.  Then substitute this value, $\alpha(\gamma,c)$,
and minimize the resulting equation respect to $\gamma$ and $c$.
However, minimizing the expectation value of the Hamiltonian when
$\gamma$ and $c$ are held fixed is equivalent to diagonalizing
the $2\times 2$-matrix
\be
   H_{2}(0,\gamma) = \left( \begin{array}{*{3}{c@{\quad}}}
    \bra{0_\gamma} H(\gamma,c) \ket{0_\gamma} & \bra{0_\gamma} H(\gamma,c) \ket{1_\gamma} \nonumber\\
     & \\
    \bra{1_\gamma} H(\gamma,c) \ket{0_\gamma} & \bra{1_\gamma} H(\gamma,c) \ket{1_\gamma}\nonumber\\
    \end{array}
    \right),
\label{Hvar}
\ee
so it is simple to show that the lowest eigenvalue of this matrix is ${\cal E}(\gamma,c)$; {\it i.e.},
\be
    {\cal E}(\gamma,c) = \frac{\bra{0_\gamma} H(\gamma,c) \ket{0_\gamma} +
    \bra{1_\gamma} H(\gamma,c) \ket{1_\gamma}}{2} -
    \sqrt{\left(\frac{\bra{1_\gamma} H(\gamma,c) \ket{1_\gamma} -
    \bra{0_\gamma} H(\gamma,c) \ket{0_\gamma}}{2}\right)^2 + \bra{0_\gamma} H(\gamma,c) \ket{1_\gamma}^2\, }
.
\ee
The eigenstate corresponding to this eigenvalue has the form
\be
    \ket{\psi_\gamma(c)} = \frac{1}{\sqrt{1+ x(c)^2}} \ket{0_\gamma} - \frac{x(c)}{\sqrt{1+x(c)^2}}
    \ket{1_\gamma} ,
\ee
where
\be
    x(c) = \frac{\sqrt{(\bra{0_\gamma} H(\gamma,c) \ket{0_\gamma}
            -\bra{1_\gamma} H(\gamma,c) \ket{1_\gamma})^2
            + 4\,(\bra{0_\gamma} H(\gamma,c) \ket{1_\gamma})^2}
            + (\bra{0_\gamma} H(\gamma,c) \ket{0_\gamma}-\bra{1_\gamma} H(\gamma,c) \ket{1_\gamma})}{2\,
            \bra{0_\gamma} H(\gamma,c) \ket{1_\gamma}} .
\ee
Note that $x(c)$ is an odd function of $c$.
The next step is to minimize ${\cal E}(\gamma,c)$ with respect to $\gamma$ and $c$.
Fig.~\ref{shiftedgaussone} shows what a typical plot looks like for finite values of $f^2$.
For some values of $\lambda$ and $f$ one can discern a secondary minimum near $c = 0$, but
it always lies higher than the minima corresponding to non-zero values of $c$.

It may seem surprising that although $c_{\rm min} \rightarrow
0 $ as $f \rightarrow 0$, it doesn't vanish at $f = 0$.  Neither does
the expectation value of the operator $x$, although it is much, much
smaller than $c_{\rm min}$.  This is because the energy is improved
by moving the gaussian slightly off $x=0$ and widening the symmetric wavefunction, admixing
a small amount of the $N=2$ state.  Note that if we rewrite the Hamiltonian,
introducing a factor $V$ as follows,
\be
    H = \frac{1}{2V}\, p^2 + \frac{V}{6}\,\lambda\, \left( x^2 - f^2 \right)^2 ,
\label{doublewellV}
\ee
and let $V \rightarrow \infty$, then $c_{\rm min} \rightarrow 0$ as
$f \rightarrow 0$.
This happens because $V$ acts as a classical {\it mass\/} for the oscillator and one expects that
for infinite mass, the wavefunction of the oscillator should become a narrower and narrower
and lie at the classical minimum of the potential.

Returning to the case $V=1$ and non-vanishing $f$, the fact that the two variational wavefunctions
$\ket{\gamma, +c_{\rm min}}$ and $\ket{\gamma, -c_{\rm min}}$ have the same energy means
that no matter how small the overlap $\bra{\gamma, -c_{\rm min}} H \ket{\gamma, +c_{\rm min}}$
they will mix maximally.  Thus it is possible to do better by considering the symmetric and
anti-symmetric wavefunctions
\ba
   \ket{\psi_{\rm even}} &=& e^{ i c p } \left( \cos(\alpha)\, \ket{0_\gamma} + \sin(\alpha)\, \ket{1_\gamma} \right)
   + e^{ -i c p} \left( \cos(\alpha)\, \ket{0_\gamma} - \sin(\alpha)\, \ket{1_\gamma} \right) \nonumber\\
   \ket{\psi_{\rm odd}} &=& e^{ i c p} \left(\cos(\alpha)\, \ket{0_\gamma} + \sin(\alpha)\, \ket{1_\gamma} \right)
   - e^{ -i c p} \left( \cos(\alpha)\, \ket{0_\gamma} - \sin(\alpha)\, \ket{1_\gamma} \right) ,
\ea
and computing the expectation value of the Hamiltonian in each of these states.
The formulas needed to carry out this computation appear in Appendix B.

The most important results are:
\begin{enumerate}
\item
For large $f^2$ the energy splitting between the even and
odd states is, as expected, exponentially small;
\item
This splitting grows as
$f^2$ tends towards zero and, in the limit of $f^2 = 0$, becomes of order unity.
\end{enumerate}

These results say that while for large $f$ it is appropriate to
think of the splitting as due to tunneling between the degenerate
minima, this is not the case when $f \rightarrow 0$.
This is true even though
$c_{\rm min}$ stays finite. The reason $c_{\rm min}$ stays finite in
the limit $f \rightarrow 0$ is that a non-vanishing value of $c_{\rm
min}$ produces a state that is an admixture of $n = 2$ and higher $n $
states with the $n =0$ state.  In fact, due to this admixing this
calculation produces energies for both the ground state and first
excited state of the pure anharmonic oscillator that are equal in
accuracy to our earlier second order adaptive perturbation theory
calculation.

\subsection{Larger $n$}

Things change for larger values of $\n$, even for $f \gg 0$.
Generically, what happens is that as $\n$ grows, $c_{\rm min}$ tends to zero, but doesn't quite
get there.  Thus as for the case $\n = 0$,
there are still two degenerate minima corresponding to equal and opposite values of $c_{\rm min}$
and so one can lower the energy by forming the states
\be
   \ket{\psi_{\rm even}} = e^{ i c p} \left( \cos(\alpha)\, \ket{n_\gamma} + \sin(\alpha)\, \ket{n+1_\gamma} \right)
   + e^{ -i c p} \left( \cos(\alpha)\, \ket{n_\gamma} - \sin(\alpha)\, \ket{n+1_\gamma} \right)
\ee
and
\be
   \ket{\psi_{\rm odd}} = e^{ i c p} \left(\cos(\alpha)\, \ket{n_\gamma} + \sin(\alpha)\, \ket{n+1_\gamma} \right)
   - e^{ -i c p} \left(\cos(\alpha)\, \ket{N_\gamma} - \sin(\alpha)\, \ket{n+1_\gamma} \right) .
\ee
The result of such a computation is to show that as $\n$ grows the splitting between these states
grows.  This follows directly from a computation of the overlap
\be
    \bra{\psi_{\rm odd}} H \ket{\psi_{\rm even}},
\ee
which, using the formulas in Appendix B, can be shown to grow as a function of $\n$.
Eventually, no matter what value we assign to $f$ (so long as it is finite) there
is a value of $\n$ for which the splitting between the states $\ket{\psi_{\rm even}}$
and $\ket{\psi_{\rm odd}}$ becomes of order unity.  This is the point at which it makes
no sense to talk about tunneling between states confined to one or the other side of the
potential barrier.

\section{Tunneling Between Asymmetric Minima}

\begin{figure}[hb]
\begin{center}
\leavevmode
\psfig{file=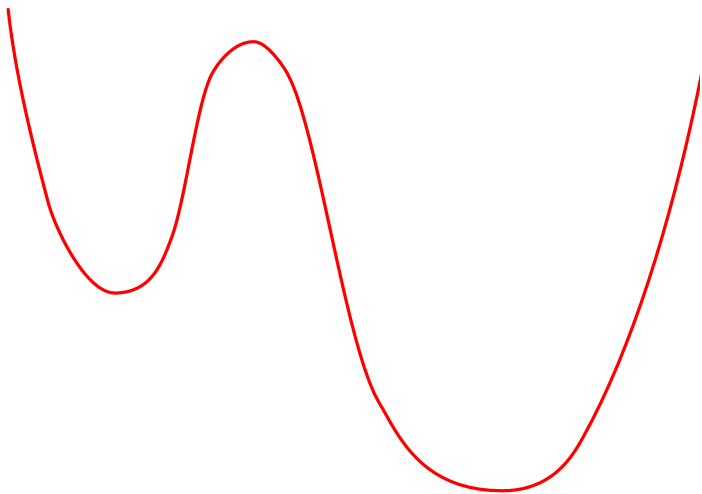,width=4in}
\end{center}
\caption{Asymmetric potential well.}
\label{asympotential}
\end{figure}

In the preceding section I showed how to use adaptive perturbation
theory to compute the tunneling between degenerate minima in a
symmetric potential.  Clearly, while the adaptive perturbation
theory discussion is a rather simple way to get accurate results, it
is by no means the only method that can be used.  However, if one
wishes to consider the tunneling in a very asymmetric potential (see
Fig.~\ref{asympotential}) then the familiar techniques are much more
difficult to apply.  In particular, consider the problem
of computing the time evolution of an
initial state corresponding to the lowest state concentrated in the
left hand well.  I know of no
simple techniques which allow one to compute this time evolution
accurately.  In this section I show how to use the
techniques of adaptive perturbation theory, plus one additional trick,
to obtain a simple picture of how things evolve in time.

From the preceding discussion, it should be clear that given a
specific form for the potential shown in Fig.~\ref{asympotential},
we can use the adaptive trick of rewriting the problem in terms of
shifted Fock states.  This allows us to approximately compute
the lowest lying state in the left well and a finite but sufficiently
large number of $n$-particle states in the right well.
These computations, augmented
by perturbation theory, take care of computing the effects which
have nothing to do with the tunneling.

Tunneling contributions are obtained by computing the transition matrix
elements between the lowest lying state shifted to the left of the
origin and all the states corresponding to a shift to the right of
the origin.  These matrix elements are computed as in the case of the
symmetric potential discussed in the preceding section. Assuming that
this step has been carried out to the desired degree of accuracy, and
that the minima are widely separated so that the transition matrix
elements are small, the next step is to compute the time evolution of
the state on the left when the tunneling is taken into account.

\subsection{The Simplified Problem}

At this point we have effectively mapped the original problem into a
simpler one; namely, that of a Hamiltonian $H$ which can be
decomposed into $ H = H_0 + V$, where $H_0$ has a discrete set of
eigenstates $\ket{\psi_j}$ with energies $E_j$, $ 1 < j < N$,
(representing the levels in the right hand well) and a single state
$\ket{\psi_0}$, with energy $E_0$ (representing the lowest lying
state of the left hand well).  Moreover, given the potential in
Fig.~\ref{asympotential}, it will be assumed that $E_0$ lies
somewhere in the range of the $E_j$, {\it i.e.}, $E_1 < E_0 < E_N$,
and that it is not degenerate with any of the $E_j$'s.  This is an
assumption I will relax at a later point in this discussion.  Note
that in this form the problem is essentially equivalent to the
Wigner-Weisskopff\cite{WW} problem of a discrete state decaying to a
continuum, except that now it is a discrete state decaying to a
large number of discrete states. Since I have assumed that adaptive
perturbation theory was used to determine the $E_j$'s and $E_0$, the
perturbation $V$ will be obtained by computing transition matrix
elements between the state $\ket{\psi_0}$ and the states
$\ket{\psi_j}$.  In other words the only non-vanishing matrix
elements of $V$ will be $V_{0j} = \bra{\psi_0} V \ket{\psi_j}$ for $
1 \le j \le N$.

I will now discuss how to compute the probability of finding the system
in the state $\ket{\psi_0}$ or $\ket{\psi_j}$ at any time $t$,
assuming it is in the state $\ket{\psi_0}$ at time $t=0$.  This is, of
course, equivalent to computing $\bra{\psi_0} e^{-i t H } \ket{\psi_0}$
or $\bra{\psi_j} e^{ -i t H} \ket{\psi_0}$.

\subsection{The Resolvent Operator}

For reasons which will become clear, the best way to calculate
$\bra{\psi_0} e^{i t H} \ket{\psi_0}$ for $t > 0$, is in terms of the resolvent operator
$1/(H-z)$.  This is because the matrix elements of the resolvent operator are directly
related to the Laplace transform of the function in question. Explicitly, consider
the Laplace transform
\be
    \psi_{00}(z) = \int_{0}^{\infty}\,dt\, e^{-z t}\, \bra{\psi_j} e^{ -i t H} \ket{\psi_0} ,
\label{Laplace}
\ee
where  $z = c + i \omega$ for fixed $c > 0$ and $-\infty \le \omega \le \infty$.
Because, in our simplified problem, the eigenvalues of $H$
are real and finite in number, this function is obviously
well defined for $z$ in the right-half complex $z$-plane, and for $t \ge 0$ the
inverse transform is given by the contour integral
\be
    \bra{\psi_j} e^{ -i t H} \ket{\psi_0} = \frac{1}{2\pi i} \int_{c-i \infty}^{c+\infty} \psi_{00}(z) e^{z t} dz .
\label{Laplaceinv}
\ee
The function is defined in the full complex plane by analytic continuation.

Carrying out the integration in Eq.~\ref{Laplace} yields
\be
    \psi_{00}(z) = -i  \bra{\psi} \frac{1}{H - i z} \ket{\psi} .
\ee
Since the resolvent operator has poles onlyl at the discrete eigenvalues of
the operator $H$, it is convenient to substitute $z \rightarrow i z$ and redefine the
contour in Eq.~\ref{Laplaceinv} to obtain
\be
     \bra{\psi} e^{i t H} \ket{\psi} = -\int_{\infty + i c}^{-\infty + i c}\,dz\,
            \bra{\psi} \frac{1}{H - z} \ket{\psi} e^{-i t z}
\ee
where we have assumed $t \ge 0$.  Obviously, the same thing can be done for two different
states; {\it i.e.},
\be
     \bra{\chi} e^{i t H} \ket{\psi} = -\int_{\infty + i c}^{-\infty + i c}\,dz\,
            \bra{\chi} \frac{1}{H - z} \ket{\psi} e^{-i t z}
\ee

The reason for focusing on the resolvent operator is
that it will be easy to compute the
relevant matrix elements of the resolvent operator exactly.  This is because the
resolvent operator satisfies the equation
\be
    \frac{1}{H - z} = \frac{1}{H_0 - z} - \frac{1}{H_0 - z}\,V\,\frac{1}{H - z} ,
\ee
and therefore the matrix element of interest satisfies the corresponding equation
\be
    \bra{\psi}\frac{1}{H - z}\ket{\psi} = \bra{\psi}\frac{1}{H_0 - z}\ket{\psi}
            - \bra{\psi} \frac{1}{H_0 - z}\,V\,\frac{1}{H - z} \ket{\psi} ,
\ee
which can be solved by iteration to obtain the formal solution
\be
    \bra{\psi} \frac{1}{H - z} \ket{\psi} = \bra{\psi} \frac{1}{H_0 - z} \sum_{n=0}^{\infty}\,(-V\,\frac{1}{H_0 - z})^n
            \ket{\psi} .
\label{integeq}
\ee

For our problem, this sum can be evaluated in closed form
because the state $\ket{\psi_0}$ is an eigenstate of $H_0$ with
eigenvalue $E_0$ and the only non-vanishing matrix elements
of $V$ are of the form $V_{0j} = \bra{\psi_0} V \ket{\psi_n}$.
From this it follows that the sum in Eq.~\ref{integeq} reduces
to a geometric series, which is easily summed to give
\be
    \bra{\psi_0} \frac{1}{H - z} \ket{\psi_0} = \frac{1}{E_0 - z} \sum_{p=0}^{\infty}
\left(\sum_j\,\frac{\bra{\psi_0} V \ket{\psi_j}\bra{\psi_j} V \ket{\psi_0}}{(E_0 - z)\,(E_j - z)}\right)^p
= \frac{1}{E_0 - z - \sum_j\,\frac{\vert \bra{\psi_0} V \ket{\psi_j}\vert|^2}{E_j - z}} .
\label{exactresum}
\ee

This can be cast into a more useful form by multiplying both the numerator and denominator
of the expression by $\prod_j (E_j - z)$ and rewriting it as a rational polynomial.
Then the poles of the resolvent operator correspond to the
zeros of the denominator and so we can rewrite the resulting expression as
\be
    \bra{\psi_0} \frac{1}{H-z}  \ket{\psi_0} =
    \frac{1}{E_0 - z - \sum_j \frac{|V_{0j}|^2}{E_j - z}}
 = \frac{\prod_{j=1}^{N} (E_j -z)}{\prod_{j=0}^N (\bar{E}_j -z)}.
\label{integeqone}
\ee
Note that here the eigenvalues of $H$ have been denoted as  $\bar{E}_j$.
It will be convenient in what follows to rewrite the $\bar{E}_j$'s in
terms of their difference from the corresponding eigenvalues of $H_0$; {\it i.e.},
\be
    \bar{E}_j = E_j + \delta_j .
\ee
It is the quantities $\delta_j$ which need to be determined.

Eq.~\ref{integeqone} is exact, independent of the size of $V$. However, since the matrix
elements of $V$ are assumed to be quite small, it is useful to rewrite it in a way that allows
the $\delta_j$'s to be computed perturbatively.
To do this, first rewrite the second term in Eq.~\ref{integeqone} as
\ba
\prod_{j=0}^N ( \bar{E_j} - z ) &=& \prod_{j=0}^N (E_j - z)\, -\, |V_{01}|^2 \prod_{j=2}^N (E_j - z)
\nonumber\\
&-& |V_{02}|^2 (E_1 - z) \prod_{j=2}^N (E_j - z)\, - \,\ldots
- |V_{0p}|^2 \prod_{j=1}^{p-1} (E_j - z)\, \prod_{l=p+1}^N (E_l - z) \nonumber\\
&-& \ldots
- |V_{0N}|^2 \prod_{j=1}^{N-1} ( E_j -z) .
\ea
From this we see that $|V_{0j}|^2 \ll 1 $ implies  $\delta_j \approx |V_{0j}|^2$.
Thus, for example, setting $j=0$ and choosing $z= E_0 + \delta_0$, the equation becomes
\ba
0 &=& -\delta_0 \,\prod_{j=1}^N (E_j-E_0) - |V_{01}|^2\,\prod_{j=2}^N - |V_{02}|^2\,(E_1-E_0)\,
\prod_{j=3}^N \nonumber\\
&-& \, \ldots \, - \, |V_{0p}|^2\,\prod_{j=1}^{p-1} (E_j - E_0) \,\prod_{l=p+1}^N (E_l - E_0)
-\,\ldots\, -|V_{0N}|^2\,\prod_{j=1}^{N-1} (E_j - E_0),
\ea
or in other words,
\be
    \delta_0 = - \sum_{j=1}^N \frac{|V_{0j}|^2}{E_j-E_0} .
\ee

The other energy shifts are obtained in a similar manner.  For example, if we set
$z= \bar{E_1} = E_1 + \delta_1$ then, to lowest order in $\delta_1$  or $|V_{0j}|^2$,
the equation becomes
\be
   0 =  (E_0 - E_1)\, (-\delta_1) \,\prod_{j=2}^N\,( E_j - E_1).
   - |V_{01}|^2\,\prod_{j=2}^N (E_j - E_1)
\ee
This is true because the other terms take the form
\be
   - |V_{02}|^2 \,(-\delta_1) \prod_{j=3}^N(E_j - E_1)\, - \,\ldots\, - \,
    |V_{0m}|^2 \,(-\delta_1) \prod_{j=2}^{m-1} (E_j - E_1) \prod_{l=m+1}^N (E_j - E_1),
\ee
which are all of second order in the perturbation.
Thus, finally, we have
\be
    \delta_1 =  \frac{|V_{01}|^2}{E_0-E_1}.
\ee
Similarly, we for all other values of $j$ we obtain
\be
    \delta_j = - \frac{|V_{0j}|^2}{E_0-E_j} .
\ee
(Note, I have implicitly assumed that the spacing between the levels $E_j$ is larger than
the typical value of $|V_{0j}|^2$ and that $E_0$ is not degenerate with any of the $E_j$'s.
The second restriction is easy to remove, since if $E_0 \approx E_j$ for some $j$, then
computing $\delta_0$ and $\delta_j$ just requires solving a quadratic
equation for its two roots.  The other $\delta_j$'s remain unchanged.  If however
the spectrum of the $E_j$'s becomes dense it is necessary to treat the problem of computing the levels
which lie within a distance of order $|V_{0j}|^2$ of $E_0$ more carefully.)

These results are half of what is needed to reconstruct the full time dependence of
$\bra{\psi_0} e^{ i t H} \ket{\psi_0}$.  It follows from Eq.~\ref{Laplaceinv} that since
the matrix element of the resolvent operator is a sum of simple poles, we can get the
full time dependence of the matrix element by closing the contour.  This would give the
time dependent behavior as a sum of exponentials multiplied by the residue of the
corresponding pole.
Fortunately, given the previous discussion, it is easy to compute the residues of the poles.
To compute these residues we begin by rewriting Eq.~\ref{integeqone} as
\be
    \bra{\psi_0} \frac{1}{H-z}  \ket{\psi_0} =
    \frac{1}{E_0 - z - \sum_j \frac{|V_{0j}|^2}{E_j - z}}
 = \frac{\prod_{j=1}^{N} (E_j -z)}{\prod_{j=0}^N (\bar{E}_j -z)} =
 \frac{\prod_{j=1}^N (\bar{E_j} - z - \delta_j)}{\prod_{s=0}^N (\bar{E_s}-z)},
\ee
and then expand the result to first order in $\delta_j$.  This yields the equation
\be
 \bra{\psi_0} \frac{1}{H-z}  \ket{\psi_0} =
\frac{1}{\bar{E_0}-z}\left( 1  - \sum_{j=1}^N \frac{\delta_j}{\bar{E_j}-z} \right).
\label{resolveone}
\ee
Now observe that since Eq.~\ref{resolveone} explicitly shows that there is a pole at every
value $z = \bar{E_j}$ it becomes a simple matter to compute the residues to the
same order in the perturbation.  Letting $z \rightarrow E_0$ yields the result
\be
    r_0 = - \left( 1 - \sum_{j=1}^N \frac{\delta_j}{\bar{E_j} - \bar{E_0}} \right),
\ee
and, similarly, letting $z \rightarrow \bar{E_j}$ implies
\be
    r_j = -\frac{\delta_j}{\bar{E_j} - \bar{E_0}}.
\ee

Combining these results, we see that the amplitude for finding the particle in the state
$\ket{\psi_0}$, at any time $t$, is given by
\be
    \bra{\psi_0}\,e^{i\,H\,t}\, \ket{\psi_0} = \sum_{j=0}^N - r_j e^{-i \bar{E_j}\,t}.
\label{tdep}
\ee
It follows from evaluating Eq.~\ref{tdep} at $t=0$ that
\be
    r_0 + \sum_{j=1}^N r_j = -1,
\ee
QED.

To get the amplitude for making a transition to state $\ket{\psi_m}$, we have to compute
\be
    \bra{\psi_m} \frac{1}{H - z} \ket{\psi_0}.
\ee
However, since $V$ only causes a transition from $\ket{\psi_m}$ to $\ket{\psi_0}$,
it follows from the integral equation
\be
    \frac{1}{H-z} = \frac{1}{H_0 -z } - \frac{1}{H_0 -z}\,V\,\frac{1}{H-z}
\ee
that
\be
\bra{\psi_m} \frac{1}{H-z}\ket{\psi_0} = \frac{V_{m0}}{E_m-z} \,\left(
\frac{\prod_{j=1}^N (E_j-z)}{\prod_{l=0}^N(\bar{E_l} -z )} \right).
\ee
Canceling the factors of $(E_j-z)$ we obtain
\be
\bra{\psi_m} \frac{1}{H-z}\ket{\psi_0} =
V_{m0}\, \frac{\prod_{j=1}^{m-1} (\bar{E_j}-z-\delta_j) \,\prod_{l=m+1}^N (\bar{E_l}-z-\delta_l)}
{\prod_{k=0}^N (\bar{E_k}-z)}.
\ee
Expanding this expression to first order in the $\delta$'s we obtain
\ba
    \tilde{r}_0 &=& - \frac{V_{m0}}{\bar{E_m}-\bar{E_0}} +
        \sum_{j=1}^{m-1} \frac{\delta_j}{(\bar{E_m}-\bar{E_0})\,(\bar{E_j}-\bar{E_0})}
        \sum_{l=m+1}^{N} \frac{\delta_l}{(\bar{E_m}-\bar{E_0})\,(\bar{E_l}-\bar{E_0})} \nonumber\\
    \tilde{r}_m &=& - \frac{V_{m0}}{\bar{E_0}-\bar{E_m}} +
       \sum_{j=1}^{m-1} \frac{\delta_j}{(\bar{E_0}-\bar{E_m})\,(\bar{E_j}-\bar{E_m})}
         \sum_{l=m+1}^{N} \frac{\delta_l}{(\bar{E_0}-\bar{E_l})\,(\bar{E_l}-\bar{E_m})}
         \nonumber\\
    \tilde{r}_j &=&
        \frac{\delta_j}{(\bar{E_0}-\bar{E_j})\,(\bar{E_m} - \bar{E_j})} ,
        \quad {\rm where}\ j \ne 0 {\rm and } \ j \ne m .
\ea
It is important to check that this solution gives the correct boundary condition
that at $t = 0$; {\it i.e.}, $\bra{\psi_m}\, e^{i\,H\,t} \, \ket{\psi_0} = 0$.
Clearly, this is the same
as requiring that the sum of the residues $\tilde{r_j}$ vanish.
Applying the preceding result we see that
\ba
\tilde{r}_0 + \tilde{r}_m &=& \sum_{j =1}^{m-1} \frac{\delta_j}{\bar{E_m} - \bar{E_0}}
\,\left( \frac{1}{\bar{E_j} - \bar{E_0}} -  \frac{1}{\bar{E_j}-\bar{E_m}} \right)
+ \sum_{j =m+1}^{N} \frac{\delta_j}{\bar{E_m} - \bar{E_0}}
\,\left( \frac{1}{\bar{E_j} - \bar{E_0}} -  \frac{1}{\bar{E_j}-\bar{E_m}} \right)
\nonumber\\
 &=& \sum_{j=1}^{m-1} \frac{-\delta_j}{(\bar{E_j}-\bar{E_0})\,(\bar{E_j}-\bar{E_m})}
+ \sum_{l=m+1}^N \frac{-\delta_l}{(\bar{E_l}-\bar{E_0})\,(\bar{E_l}-\bar{E_m})},
\ea
and therefore,
\be
\tilde{r}_0 + \tilde{r}_m + \sum_{j=1}^{m-1} \tilde{r}_j + \sum_{l=m+1}^N \tilde{r}_l = 0,
\ee
which is the desired result.

Given these formulas, we can write the time-dependent state $\ket{\psi_0(t)}$ as
\be
    \ket{\psi_0(t)} = e^{ i t H} \ket{\psi_0} =
           \sum_{m=0}^N \, \bra{\psi_m} e^{i t H} \ket{\psi_0} \ket{\psi_m}.
\ee
It is easy to see that in the case where the eigenvalues of $H_0$ are widely spaced
on the scale of $|V_{0j}|^2$, the largest residues, and therefore the largest contribution
to $\ket{\psi_0(t)}$, comes from $\ket{\psi_0}$ and the state $\ket{\psi_m}$ for which
$E_m-E_0$ takes the smallest value.  However taking the limit in which the levels
in the right well become dense, as would be the case if the right hand minimum is very
flat and very wide, cause things to change.
While the formula Eq.~\ref{exactresum} remains exact the sum over levels is best approximated
by an integral over a continuous variable $E$; {\it i.e.},
\be
    \bra{\psi_0} \frac{1}{H - z} \ket{\psi_0} = \frac{1}{E_0 - z} \sum_{p=0}^{\infty}
\left(\int_{E_{min}}^{E_{max}}\,\frac{\bra{\psi_0} V \ket{\psi(E)}\bra{\psi(E)} V \ket{\psi_0}}{(E_0 - z)\,(E_j - z)}\right)^p
= \frac{1}{E_0 - z - \int_{E_{min}}^{E_{max}}\,\frac{\vert \bra{\psi_0} V \ket{\psi(E)}\vert|^2}{E - z}} .
\ee
Since in this case $E_0$ lies in the region of integration, we see immediately from the fact that
the integral develops an imaginary part for $E_{min} \le z \le E_{max}$,
that the resolvent operator no longer has a pole at real
$z \approx E_0$.  Instead, the zero of the denominator, and thus the pole of the resolvent operator,
moves off to a complex value of $z$ on the second sheet of the function.  In order to compute
the time dependence of the state one has to deform the contour for the inverse transform and
pick up this pole.  This will give the usual exponential decay of the state $\ket{\psi_0}$ plus a background
integral that does not decay exponentially in $t$ for large $t$.

\section{A Look Ahead to Field Theory}

Before concluding, it is worth taking a quick look ahead to the case of an
interacting scalar field theory, to see how the techniques of
adaptive perturbation theory can be used to non-perturbatively derive the structure
of renormalization.  In this section I focus on a simple calculation that
gives a non-perturbative approximation to mass renormalization and I will leave the discussion of
wavefunction and coupling constant renormalization to the next paper.

To simplify the discussion I will limit myself to the case of a self-interacting
scalar field, with a Hamiltonian of the form
\be
    H = \int d^nx \left[ \frac{1}{2} \Pi_{\phi}(x)^2 + \frac{1}{2} \nabla \phi(x)\cdot \nabla\phi(x)
               + \frac{\lambda}{6} \phi(x)^4 \right] .
\ee
The spatial dimension $n$ is arbitrary.  In what follows I assume that everything
is defined in a finite volume, $V$, so that momenta are discrete.  This isn't necessary, but it will
make all of the manipulations which follow well-defined.  Given this assumption,
the Fourier transforms of the field and its conjugate momentum are defined to be
\be
    \phi(\vec{k}) = \frac{1}{\sqrt{V}} \int d^nx e^{-i \vec{k}\cdot\vec{x} } \phi(x) \quad ; \quad
    \Pi_{\phi}(\vec{k}) = \frac{1}{\sqrt{V}} \int d^nx e^{-i \vec{k}\cdot\vec{x} } \Pi_{\phi}(x).
\ee
In terms of these operators the Hamiltonian can be rewritten as
\be
    H =  \sum_{\vec{k}} \left[ \frac{\Pi_\phi(-\vec{k}) \Pi_\phi(\vec{k})}{2} +\frac {(
\vec{k}\cdot\vec{k} + m^2)}{2}
    \,\phi(\vec{k})\,\phi(-\vec{k})\right]
    + \frac{\lambda}{6} \frac{1}{V} \sum_{\vec{k}_i} \phi(\vec{k}_1)\,\phi(\vec{k}_2)
\phi(\vec{k}_3)\,\phi(\vec{k}_4) \delta(\sum_{i=1}^4 \vec{k}_i).
\ee

At this point I set up the adaptive perturbation theory calculation by defining
$\gamma(k)$-dependent annihilation and creations operators as follows:
\be
    \phi(\vec{k}) = i\,\sqrt{\frac{\gamma(\vec{k})}{2}} ( {\cal A}(-\vec{k})^{\dag} - {\cal A}(\vec{k}) ) \quad ;\quad
    \Pi(\vec{k}) = \frac{( {\cal A}(-\vec{k})^{\dag} + {\cal A}(\vec{k}) )}{\sqrt{2\,\gamma(\vec{k})}},
\ee
and define the associated {\it vacuum\/} state for this choice of $\gamma(\vec{k})$'s,
\be
    \ket{vac} = \prod_k \ket{0_{\gamma(k)}},
\ee
by the condition that it be annihilated by all the ${\cal A}(\vec{k})$'s.

Adaptive perturbation theory around this vacuum state is now
determined by minimizing the vacuum expectation value of the
Hamiltonian with respect to all of the $\gamma(\vec{k})$'s.  It
follows directly from these definitions that the function to be
minimized is
\be
\bra{vac} \, H\, \ket{vac} =  \sum_k \left[ \frac{\gamma(\vec{k})}{4}
+ \frac{( \vec{k}\cdot\vec{k} + m^2 )}{4\,\gamma(\vec{k})} \right]
+ \frac{\lambda}{4 V} \left[  \sum_{\vec{k}} \frac{1}{\gamma(\vec{k})} \right]^2.
\label{vacenergy}
\ee
If the range of the momenta appearing in these sums are
unrestricted, these expressions diverge, as happens in all field
theories.  The usual way this problem is dealt with in the context
of ordinary perturbation theory is to regulate the integrals and then add
counterterms to the Lagrangian to cancel the divergences.  Since I
wish to discuss this theory non-perturbatively, I will adopt a
different strategy.  I will render the theory well defined by
restricting the operators $\phi(\vec{k})$ and $\Pi_\phi(\vec{k})$ to
be finite in number.  This can be accomplished
in a variety of ways.  One is to put the theory on a
lattice by defining the momentum components to be $ k_\mu(j) = 2\pi
j/L_\mu a $, where $L_\mu$ is the size of the lattice in the $\mu$
direction, $a$ is the lattice spacing and $j$ is an integer running
from $-L_\mu/a \le j \le L_\mu/a$.  Another
way to accomplish the same thing is to keep only those operators
$\phi(\vec{k})$ and $\Pi_\phi(\vec{k})$ for which $\vec{k}\cdot\vec{k}
\le \Lambda$. Since nothing I will say from this point on requires
that I specify a particular scheme, I avoid making any detailed
assumptions.

Minimizing Eq.~\ref{vacenergy} with respect to each $\gamma(\vec{k})$ yields
\be
    \frac{1}{4} - \frac{( \vec{k}\cdot\vec{k} + m^2 ) }{4\,\gamma(k)^2}
- \frac{\lambda}{2 \gamma(\vec{k})^2} \left[\frac{1}{V} \sum_{\vec{k}'} \frac{1}{\gamma(\vec{k}')} \right] = 0,
\ee
which can be rearranged to give
\be
    \gamma(k)^2 = \vec{k}\cdot\vec{k} + m^2
    + 2\,\lambda\,\left[\frac{1}{V} \sum_{\vec{k}'} \frac{1}{\gamma(\vec{k}')} \right].
\label{generick}
\ee
Setting $\vec{k}=0$ we obtain
\be
    \gamma(0)^2 = m^2 + 2\,\lambda\,\left[ \frac{1}{V} \sum_{\vec{k}'} \frac{1}{\gamma(\vec{k}')} \right] ,
\ee
which can be substituted into Eq.~\ref{generick} to give
\be
    \gamma(\vec{k})^2 = \vec{k}\cdot\vec{k} + \gamma(0)^2 .
\ee
Substituting this expression into the equation for
$\gamma(0)$ yields the non-perturbative equation
\be
    \gamma(0)^2 = m^2
    + 2\,\lambda\,\frac{1}{V} \sum_{\vec{k}'} \frac{1}{\sqrt{\vec{k}'\cdot\vec{k}' + \gamma(0)^2}},
\ee
which is reminiscent of the Nambu Jona-Lasinio equation.  This
equation is guaranteed to have a solution since the formula for the vacuum
energy clearly diverges both at $\gamma(0) =0$ and $\gamma(0) =
\infty$ and is finite in between. At this point we can take the
infinite volume limit and replace the sum over $\vec{k}$'s by an
integral up to some cut-off $\Lambda$.  In this case the equation to
be solved becomes
\be
    \gamma(0)^2 - m^2 =
     2\,\lambda\,\frac{2^p}{(2\pi)^p} \int_0^{\Lambda} k^{p-1} dk
    \frac{1}{\sqrt{\vec{k}\cdot\vec{k} + \gamma(0)^2}}.
\label{NJL}
\ee
The simplest way to understand the general properties of the
solution to this equation is to graph the left and right hand sides
as a function of $\gamma(0)$ and see where they cross. Clearly, the
graph of the left hand side of Eq.~\ref{NJL} is a parabola that
intersects the vertical axis at $-m^2$.  The graph of the right hand
side of the equation starts at some large (or in 1+1-dimension
infinite) value and then drops monotonically to zero as $\gamma(0)
\rightarrow \infty$.  Generically, since for finite $\gamma(0) \ll
\Lambda$, the integral is proportional to $\Lambda^{p-1}$ for $p > 1$,
and $\ln(\Lambda)$ for $p=1$, the point where the parabola crosses the
graph of the integral will be for large $\gamma(0)$.  The only way to
have the graphs cross for a value of $\gamma(0)$ that is small with
respect to $\Lambda$ is to make $m^2$ large and negative.
In fact, one should really think of Eq.~\ref{NJL} not as an equation
for $\gamma(0)$, but rather as an equation for $m^2$; {\it i.e.}, rewrite
it as
\be
    m^2 = \gamma(0)^2
    - 2\,\lambda\,\frac{2^p}{(2\pi)^p} \int_0^{\Lambda} k^{p-1} dk
    \frac{1}{\sqrt{\vec{k}\cdot\vec{k} + \gamma(0)^2}}
\ee
and then arbitrarily pick a value of $\gamma(0)$ and use this equation
to determine for what value of $m^2$ the chosen value of $\gamma(0)$
will minimize the ground state energy density.
This is nothing but a non-perturbative way of determining
the leading mass renormalization counter term.

While this discussion is amusing, it does nothing to show what one
has to do to similarly understand wavefunction and coupling
constant renormalization in a non-perturbative context.  That
discussion will be the subject of the second paper in this series.

\section{Reprise}

It was my purpose in this paper to introduce the notion of {\it adaptive
perturbation theory\/} as a way of perturbatively computing quantities
that heretofore couldn't be computed in any simple way.  I
focused on problems in ordinary quantum mechanics such as the pure
anharmonic oscillator and the problem of computing the eigenstates
of generic double-well problems in order to exhibit the versatility of the
technique.  Finally, I discussed the simplest way to begin to apply
these techniques to a field theory in order to obtain a better behaved
perturbation expansion independent of the size of the interaction term.
In addition, I tried to indicate how the same approach would allow one
to non-perturbatively approach the computation of renormalized quantities.
In the next paper I will show how to extend these techniques to
non-perturbatively capture the full structure of renormalization.
I will also show how something related to the running coupling constant appears and
explain why this approach essentially forces one to discuss the computation
of scattering amplitudes in a way that is intimately related to the
approach used in the parton model.

Obviously there are many items which I have not touched upon and
which require closer analysis.  One particularly interesting
unsolved problem, which is relevant to the analysis of a compact gauge
theory, has to do with a problem in which the potential has an
infinite number of degenerate minima. When these minima are widely
separated by barriers of sufficient height then one can do a good
job simply by using the adaptive technique to compute states
localized in any one minimum and then summing over appropriate
superpositions of these states to obtain Bloch wave function.
However when there is significant tunneling between nearby minima
one has to better understand, as we did in the simple tunneling
problem, how to include single-well excited states in the trial
wavefunction so as to avoid double counting in the perturbative
expansion.

The key point I would like the reader to take away from this discussion is
that, in an interacting theory, the perturbation theory one uses to
compute an interesting quantity must be adapted to the problem at hand.
Although a variational computation based on energy considerations is the
tool used to define the appropriate parameters to be used when
dividing the Hamiltonian up into an {\it unperturbed \/} part and
and an {\it interaction term\/}, it is the perturbation theory based
upon this division which is crucial to computing quantities other
than the energy of a state.  This observation takes us
beyond the more usual variational methods which cannot be usefully applied
to compute matrix elements of general operators, or scattering amplitudes
in a field theory.

\section{Acknowledgements}

I would like to thank Carl Bender for useful communications and in
particular for pointing out to me the existence of the papers by
Halliday and Suranyi\cite{Halliday:1979vn}.

\section*{APPENDIX A. SOME OPERATOR IDENTITES}

This appendix is devoted to reminding the reader how to derive
normal ordering identities that are useful for carrying out
the computations I have already described, as well as the
computations in Appendix B.

\parskip 10pt
\noindent{\bf Theorem:}
\be
    e^{\alpha (\Adag + \A)} = e^{\alpha^2/2}\,e^{\alpha \Adag} \,e^{\alpha \A}
\label{normalorderx}
\ee

\noindent{\bf Proof:} \quad
Assume
\be
    e^{\alpha (\Adag + \A)} = e^{f(\alpha)}\,e^{g(\alpha) \Adag}\,e^{h(\alpha)\A}
\ee
and differentiate with respect to $\alpha$ to obtain
\be
    (\Adag + \A)\,e^{\alpha (\Adag + \A)} = \left[\frac{d}{d\alpha}f(\alpha) +
    \frac{d}{d\alpha}g(\alpha) \Adag + \frac{d}{d\alpha}h(\alpha)\,e^{g(\alpha) \Adag}\,e^{h(\alpha)\A}\,
    e^{-g(\alpha)\Adag} \right] \,e^{f(\alpha)}\,e^{g(\alpha) \Adag}\,e^{h(\alpha)\A}.
\label{identityone}
\ee
Successive differentiation of the term $e^{g(\alpha) \Adag}\,e^{h(\alpha)\A}\,e^{-g(\alpha)\Adag}$
with respect to $\alpha$ gives the identity
\be
e^{g(\alpha) \Adag}\,e^{h(\alpha)\A}\,e^{-g(\alpha)\Adag} = \sum_n \frac{(g(\alpha)^n}{n!} [ \Adag,[\Adag,[
\ldots,[\Adag,\A]]]\ldots ]_n .
\ee
Observing that $[\Adag,\A]=-1$ allows us to rewrite this as
\be
e^{g(\alpha) \Adag}\,e^{h(\alpha)\A}\,e^{-g(\alpha)\Adag} = \A - g(\alpha).
\ee
Substitute this into Eq.~\ref{identityone} and cancel the factor of $e^{\alpha (\Adag + \A)}$
against the term $e^{f(\alpha)}\,e^{g(\alpha) \Adag}\,e^{h(\alpha)\A}$ on the right
hand side of the equation to obtain
\be
    (\Adag + \A) = \frac{d}{d\alpha}f(\alpha) +
    \frac{d}{d\alpha}g(\alpha) \Adag + \frac{d}{d\alpha}h(\alpha)\,\left(\A - g(\alpha)\right).
\ee
Equating coefficients of $\Adag$, $\A$ and the unit operator yields the following differential
equations:
\be
    \frac{d}{d\alpha}g(\alpha) = 1 \quad ; \quad
    \frac{d}{d\alpha}h(\alpha) = 1 \quad ; \quad
    \frac{d}{d\alpha}f(\alpha) = g(\alpha)\,\frac{d}{d\alpha}h(\alpha)
\ee
The initial conditions $f(0)=0$,$g(0)=0$ and $h(0)=0$ follow from the fact that $e^{\alpha (\Adag + \A)}$
is the identity operator for $\alpha = 0$.  Given these boundary conditions, it follows directly that
\be
    g(\alpha) = \alpha \quad ; \quad h(\alpha) = \alpha \quad ;\quad f(\alpha) = \frac{\alpha^2}{2},
\ee
which is what we wished to prove.

A similar calculation proves the following result:

\parskip 10pt
\noindent{\bf Theorem:}
\be
 e^{\alpha ( \Adag - \A )} = e^{-\alpha^2/2}\,e^{\alpha \Adag} \,e^{-\alpha \A}
\label{normalorderp}
\ee

This identity is used to compute, among other things, the overlap $\bra{0_\gamma} e^{-i c p} \ket{0_\gamma}$
as follows:
\be
\bra{0_\gamma} e^{-i c p} \ket{0_\gamma} = \bra{0_\gamma} e^{c\,\sqrt{\frac{\gamma}{2}} ( \Adag - \A)}
\ket{0_\gamma} = e^{-c^2\gamma/4}\,\bra{0_\gamma}\,e^{c\,\sqrt{\frac{\gamma}{2}}\Adag}\,e^{c\,\sqrt{\frac{\gamma}{2}}\A}
\ket{0_\gamma} = e^{-c^2\gamma/4},
\ee
since, by assumption, $\A \ket{0_\gamma} = 0$

I mentioned that there was a simple expression for the state $\ket{0_\gamma}$ in terms of
the states built upon $\ket{0_{\gamma'}}$.  The explicit form of this relation is the following:

\parskip 10pt
\noindent{\bf Theorem:}
\be
    \ket{0_\gamma}= \sqrt{1 - \tanh(\beta)^2}\,e^{-\frac{1}{2}\,\tanh(\beta)\,\Adag^2} \ket{0_{\gamma'}},
\ee
where $\tanh(\beta) = (\gamma - \gamma')/(\gamma + \gamma')$.

\noindent{\bf Proof:}
Given that we introduced that $\gamma$-dependent annihilation and creation operators as
\be
    x = \frac{\Adag + \A}{\sqrt{2 \gamma}} \quad ; \quad p = i \sqrt{\frac{2}{\gamma}} (\Adag - \A)
\ee
it follows that for two different values of $\gamma$
\be
    \A = \frac{1}{2} \frac{(\gamma + \gamma')}{\sqrt{\gamma\,\gamma'}}\, A_{\gamma'}
    + \frac{1}{2} \frac{(\gamma - \gamma')}
    {\sqrt{\gamma\,\gamma'}}\,A^{\dag}_{\gamma'} .
\ee
By definition $\A \ket{0_\gamma} = 0$. Thus we have
\be
 \frac{1}{2} \frac{(\gamma + \gamma')}{\sqrt{\gamma\,\gamma'}}\, A_{\gamma'} \ket{0_\gamma}
    + \frac{1}{2} \frac{(\gamma - \gamma')}
    {\sqrt{\gamma\,\gamma'}}\,A^{\dag}_{\gamma'}\ket{0_\gamma} = 0 .
\ee
Now, consider a state of the form
\be
    e^{\lambda {A^{\dag}_{\gamma'}}^2} \ket{0_{\gamma'}}.
\ee
Requiring that $\A$ annihilates this state is equivalent to the statement
\be
 \frac{1}{2} \frac{(\gamma + \gamma')}{\sqrt{\gamma\,\gamma'}}\, A_{\gamma'} e^{\lambda\, (A^{\dag}_{\gamma'})^2} \ket{0_{\gamma'}}
    + \frac{1}{2} \frac{(\gamma - \gamma')}
    {\sqrt{\gamma\,\gamma'}}\,A^{\dag}_{\gamma'}e^{\lambda\, (A^{\dag}_{\gamma'})^2} \ket{0_{\gamma'}} = 0 .
\ee
Moving the factor $ e^{\lambda {A^{\dag}_{\gamma'}}^2} $ to the left of each term we obtain
\be
 e^{\lambda {A^{\dag}_{\gamma'}}^2} \left[
 \frac{1}{2} \frac{(\gamma + \gamma')}{\sqrt{\gamma\,\gamma'}}\, e^{-\lambda\, (A^{\dag}_{\gamma'})^2}\,A_{\gamma'}\,
 e^{\lambda\, (A^{\dag}_{\gamma'})^2} + \frac{1}{2} \frac{(\gamma - \gamma')}
    {\sqrt{\gamma\,\gamma'}}\,A^{\dag}_{\gamma'} \right]\ket{0_\gamma'} = 0 .
\ee
Once again, expanding the term $ e^{-\lambda {A^{\dag}_{\gamma'}}^2}\,A_{\gamma'}\,
 e^{\lambda {A^{\dag}_{\gamma'}}^2}$ in multiple commutators we obtain the results that $\A$
will annihilate the state if and only if
\be
 \lambda \frac{(\gamma + \gamma')}{\sqrt{\gamma\,\gamma'}} + \frac{1}{2} \frac{(\gamma - \gamma')}
    {\sqrt{\gamma\,\gamma'}}  = 0,
\ee
or, in other words, if
\be
    \lambda = \frac{1}{2}\left( \frac{\gamma - \gamma'}{\gamma + \gamma'} \right) = \tanh(\beta),
\ee
QED.

Finally, the normalization factor come from the identity
\be
    e^{\lambda\,{A^{\dag}_{\gamma'}}^2} \ket{0_{\gamma'}} =
    \sum_{n=0}^{\infty} \frac{\lambda^n}{n!} \sqrt{2n!} \ket{2n_{\gamma'}}
\ee
The norm squared of this state is given by the sum
\be
  \sum_{n=0}^{\infty} \frac{\lambda^{2n}}{{n!}^2} \,{2n}! = \frac{1}{\sqrt{1 - 4\lambda^2}},
\ee
so the normalized state is
\be
\ket{0_{\gamma}} = (1-\tanh(\beta)^2)^{1/4}\,  e^{\frac{1}{2}\tanh(\beta)\,{A^{\dag}_{\gamma'}}^2} \ket{0_{\gamma'}}.
\ee
It is worth noting that another way to derive this result is to observe that
\be
    e^{\frac{\beta}{2}( {A^{\dag}}^2 - {A}^2)}
\ee
is a unitary operator that satisfies the relation
\be
    e^{\frac{\beta}{2}( {A^{\dag}}^2 - {A}^2)} =
    e^{-\frac{1}{2} \tanh(\beta) {A^{\dag}}^2}\,e^{-\ln(\cosh(\beta))\,(A^{\dag}A +\frac{1}{2})}
    \,e^{\frac{1}{2} \tanh(\beta) {A^{\dag}}^2} .
\ee
This is proven in much the same way as are the other normal ordering relations.
Putting back the appropriate $\gamma'$ dependence in this equation and applying the
operator to the state $\ket{0_{\gamma'}}$  gives the desired result.

\section*{APPENDIX B. DETAILS OF DOUBLE WELL CALCULATION}

As we already noted, the two degenerate variational states for non-vanishing $f^2$ are defined
as:
\be
\ket{\chi(c, \gamma)} = e^{- i c p} \ket{\psi_{\gamma}(c)}
\ee
and
\be
    \ket{\chi(-c, \gamma)} = e^{ i c p} \ket{\psi_{\gamma}(-c)}
\ee
(assuming $c>0$ throughout).

Once one has found the best choice of variational
parameters, still lower energy states are obtained by forming the
even and odd parity combinations:

\be
\ket{\psi_{\rm even}(\gamma)} = \frac{1}{N_{\rm even}} \left(\, \ket{\chi(c, \gamma)}
    + \ket{\chi(-c, \gamma)} \,\right)
\ee
 and
\be
    \ket{\psi_{\rm odd}(\gamma)}
    = \frac{1}{N_{\rm odd}} \left( \,\ket{\chi(c, \gamma)} - \ket{\chi(-c, \gamma)} \,\right) ,
\ee
where $N_{\rm even}$ and $N_{\rm odd}$ are normalization factors.

To evaluate the normalization factors $N_{\rm even}$ and $N_{\rm odd}$ we need to compute
the scalar product $ \langle \chi(c, \gamma) \vert \chi(-c, \gamma) \rangle $
as well as overlap elements $\bra{ \chi(c, \gamma)} H \ket{\chi(-c, \gamma)}$. Since the diagonal scalar
products are obviously unity, the diagonal expectation values of
the Hamiltonian are simply the values previously computed by minimizing
${\cal E}(c, \gamma)$.

Expressing $p$ as a sum of annihilation and creation operators and using the identity for
computing the normal ordered exponential, Eq.~\ref{normalorderp}, allows us to rewrite
the overlap as
\be
\langle \chi(-c,\gamma) \vert \chi(c,\gamma) \rangle =
\bra{\psi_{-c}(\gamma)} e^{-2 i c p} \ket{\psi_c(\gamma)}
=  e^{-c^2 \gamma} \bra{\psi_{-c}(\gamma)} e^{
c \sqrt{2\gamma} \Adag }\, e^{-c \sqrt{2\gamma} \A } \ket{\psi_c(\gamma)}
= e^{-c^2 \gamma}
\langle \phi_{-c}(\gamma) \vert \phi_c(\gamma) \rangle ,
\ee
where the states $\bra{\phi_c(\gamma)}$ and $\ket{\phi_{-c}(\gamma)}$ are defined in terms of the
states
\be
\ket{\psi_c(\gamma)} = \frac{1}{\sqrt{1+x(c)^2}}\, \ket{\psi_0(\gamma)} -
 \frac{x(c)}{\sqrt{1+x(c)^2}}\, \ket{\psi_1(\gamma)}
\ee
and
\be
\ket{\psi_{-c}(\gamma)} = \frac{1}{\sqrt{1+x(c)^2}}\,
 \ket{\psi_0(\gamma)} + \frac{x(c)}{\sqrt{1+x(c)^2}}\, \ket{\psi_1(\gamma)}.
\ee
Note that in this formula $x(c)$ is given by the formula evaluated for positive $c$, and
the states $\ket{\phi_c(\gamma)}$ and $\ket{\phi_{-c}(\gamma)}$ are given by
\be
\ket{\phi_c(\gamma)} =  e^{-c \sqrt{2\gamma} \A }\, \ket{ \psi_c(\gamma)}
\ee
and
\be
   \ket{\phi_{-c}(\gamma)} = e^{c \sqrt{2 \gamma} \A }\, \ket{\psi_{-c}(\gamma)} .
\ee

Since the state $\ket{\psi_c(\gamma)}$ contains only $n = 0$ and $n = 1$ states,
only the first two terms in the expansion of the exponential act.
Thus it follows that
\be
\ket{\phi_c(\gamma)} =  e^{-c\sqrt{2 \gamma}\,\A} \ket{\psi_c(\gamma)} = \left(
\frac{1+ c\, \sqrt{2 \gamma}\, x(c)}{\sqrt{1 + x(c)^2}} \right)\ket{\psi_0(\gamma)}-
\frac{x(c)}{\sqrt{1+x(c)^2}} \ket{\psi_1(\gamma)}
\ee
and
\be
\ket{\phi_{-c}(\gamma)} =
e^{c \sqrt{2 \gamma}\, \A} \ket{\psi_{-c}(\gamma)}
= \left( \frac{1 + c\, \sqrt{2 \gamma}\, x(c)}{\sqrt{1+x(c)^2}} \right)\, \ket{\psi_0(\gamma)}
+ \frac{x(c)}{\sqrt{1+x(c)^2}}\, \ket{\psi_1(\gamma)},
\ee
so the overlap is
\be
\langle \chi(-c,\gamma) \vert \chi(c,\gamma) \rangle
= e^{-c^2 \gamma} \langle \phi_{-c}(\gamma) \vert \phi_c(\gamma) \rangle
= e^{-c^2 \gamma}\,\frac{\left(1+c \sqrt{2 \gamma}\, x(c) \right)^2-x(c)^2}{1+x(c)^2}.
\ee
Thus, the normalization for the state $\ket{\psi_{\rm even}(\gamma)}$ is
\be
N_{\rm even} = \sqrt{(1+ \ov)} .
\ee
Similarly
\be
N_{\rm odd} = \sqrt{2 (1- \ov)}.
\ee

Finally, to compute the formula for the expectation value of
$H$ in the state $\ket{\psi_{\rm even}(c, \gamma)}$ or $\ket{\psi_{\rm odd}(c, \gamma)}$,
we need the following identities:
\ba
e^{-i c p}\, H(p, x)\, e^{i c p} &=& H(p, x-c) \nonumber\\
e^{i c p}\, H(p, x)\, e^{-i c p} &=& H(p, x+c) \nonumber\\
e^{-i c p}\, H(p, x)\, e^{-i c p } &=& H(p, x-c)\, e^{-2 i c p}
= e^{- c^2 \gamma }\, H(p, x-c)\, e^{c \sqrt{2 \gamma} \Adag }\, e^{-c \sqrt{2 \gamma} \A} \nonumber\\
&=& e^{- c^2 \gamma}\, e^{c \sqrt{2 \gamma} \Adag}\, H(p- i c \gamma, x)\, e^{-c \sqrt{2 \gamma} \A}
\nonumber\\
e^{i c p}\, H(p, x)\, e^{ i c p} &=& H(p, x+c)\, e^{2 i c p}
= e^{-c^2 \gamma} \,H(p, x+c)\, e^{ c \sqrt{2 \gamma} \Adag}\, e^{c \sqrt{2 \gamma} \A} \nonumber\\
 &=& e^{-c^2 \gamma}\, e^{-c \sqrt{2 \gamma} \Adag}\, H(p + i c \gamma, x)\, e^{c \sqrt{2 \gamma} \A}.
\ea
We then need to evaluate these expressions between the
states $\ket{\psi_c(\gamma)}$ and $\ket{\psi_{-c}(\gamma)}$.
Clearly, the first two terms are equal to
${\cal E}(\lambda, f^2, c, \gamma)_{\rm min}$, which is the minimum value of ${\cal E}(\lambda, f^2,
c, \gamma)$ obtained by minimizing with respect to $c$ and $\gamma$, since the minima are degenerate.

Given these identities, it is straightforward to evaluate
\ba
\bra{\psi_{\rm even}} H \ket{\psi_{\rm even}}
=  \frac{1}{N_{\rm even}^2} & &\left[ \bra{\chi(c,\gamma)} H(p,x) \ket{\chi(c,\gamma)}
+ \bra{\chi(-c,\gamma)} H(p,x) \ket{\chi(-c, \gamma)}  \right. \nonumber\\
 & & \left. + \ \bra{\chi(-c,\gamma)} H(p,x) \ket{\chi(c,\gamma)}
  + \bra{\chi(c,\gamma)} H(p,x) \ket{\chi(-c,\gamma)} \right].
\ea
Clearly, the first two terms evaluate to the function ${\cal E}(\lambda, f^2, c, \gamma)$
and the remaining terms are
\be
 \bra{\chi(-c,\gamma)} H(p,x) \ket{\chi(c,\gamma)}
 = e^{-c^2 \gamma} \bra{(1+c \sqrt{2 \gamma}\A) \psi_{-c}(\gamma) }
 H(p- i c \gamma, x) \ket{( 1 - c \sqrt{2 \gamma} \A) \psi_c(\gamma) }
\ee
and
\be
\bra{\chi(c, \gamma)} H(p,x) \ket{\chi(-c, \gamma)} =
e^{-c^2 \gamma} \bra{( 1 - c \sqrt{2 \gamma}\A) \psi_c(\gamma)} | H(p + i c \gamma,x)
\ket{(1 + c \sqrt{2 \gamma}\A) \psi_{-c}(\gamma)} .
\ee

Explicit substitution of the shifted definition of $p$ in $H$ yields
\be
 H(p- i c \gamma, x) = H(p, x) - i c \gamma p - \frac{c^2}{2} \gamma^2
\ee
and
\be
H(p + i c \gamma, x) = H(p, x) + i c \gamma p - \frac{c^2}{2} \gamma^2 .
\ee
To relate the two cross-terms, it is helpful to observe that
\be
\bra{\chi(-c,\gamma)} O \ket{\chi(c,\gamma)}
= \bra{\chi(c,\gamma)} O^{\dag} \ket{\chi(-c, \gamma)}^{\ast} .
\ee
Since $i\,p$ is anti-hermitian, we get a minus sign for that term and plus signs for all the others.
Note that the expectation value of $H(p,x)$ that appears in the cross terms is
obtained from Eq.~\ref{Hvar} setting $c = 0$ and evaluating the result between the states $\bra{\phi_{-c}(\gamma)}$
and $\ket{\phi_c(\gamma)}$.

The result for the expectation value of the
Hamiltonian in the variational states $\ket{\psi_{\rm even}}$ and $\ket{\psi_{\rm odd}}$
is:

\ba
\bra{\psi_{\rm even}} H \ket{\psi_{\rm even}}  &=&
\frac{2}{N_{\rm even}^2} \left[{\cal E}(\lambda, f^2, c, \gamma)
+ \frac{e^{-c^2\,\gamma}}{1+x(c)^2}\,\left((1 + \sqrt{2 \gamma}\,c\,x(c))^2
\left(\frac{\gamma}{4} + \frac{\lambda\,f^4}{6} + \frac{\lambda}{8 \gamma^2}
- \frac{\lambda f^2}{6 \gamma}- \frac{c^2\,\gamma^2}{2}\right)  \right.\right. \nonumber\\
&+& \left.\left. 2\,x(c)\,(1 + \sqrt{2 \gamma}\,c\,x(c))\,c\,\gamma\,\sqrt{\frac{\gamma}{2}}
- x(c)^2\,\left(
\frac{5 \lambda}{8 \gamma^2} + \frac{3 \gamma}{4} - \frac{\lambda \,f^2}{2 \gamma}
+ \frac{\lambda\,f^4}{6} - \frac{c^2 \gamma^2}{2} \right) \right)\right]
\ea
and
\ba
\bra{\psi_{\rm odd}} H \ket{\psi_{\rm odd}}  &=&
\frac{2}{N_{\rm even}^2} \left[{\cal E}(\lambda, f^2, c, \gamma)
- \frac{e^{-c^2\,\gamma}}{1+x(c)^2}\,\left((1 + \sqrt{2 \gamma}\,c\,x(c))^2
\left(\frac{\gamma}{4} + \frac{\lambda\,f^4}{6} + \frac{\lambda}{8 \gamma^2}
- \frac{\lambda f^2}{6 \gamma}- \frac{c^2\,\gamma^2}{2}\right)  \right.\right. \nonumber\\
&+& \left.\left. 2\,x(c)\,(1 + \sqrt{2 \gamma}\,c\,x(c))\,c\,\gamma\,\sqrt{\frac{\gamma}{2}}
- x(c)^2\,\left(
\frac{5 \lambda}{8 \gamma^2} + \frac{3 \gamma}{4} - \frac{\lambda \,f^2}{2 \gamma}
+ \frac{\lambda\,f^4}{6} - \frac{c^2 \gamma^2}{2} \right) \right)\right]
\ea

So long as the factor $e^{-c^2\,\gamma}$ is small, the splitting between the even
and odd states is quite small.  Thus the tunneling between the left and right states is slow.
When we go to smaller values of $f$ or larger values of $n$, the tunneling rate increases
and even at $f = 0$ minimizing $\bra{\psi_{\rm even}} H \ket{\psi_{\rm even}}$ and
$\bra{\psi_{\rm odd}} H \ket{\psi_{\rm odd}}$ with respect to $\gamma$ and $c$ leads
to a non-vanishing value for $c$.  The energies obtained in this way
agree in accuracy with the second order adaptive perturbation theory
calculation for the pure anharmonic oscillator, which was done without using the shifted
wavefunction.  As expected, this shows that the effect of having a non-vanishing value
of $c$ at $f=0$ is to include the effects of the second order computation.

\clearpage

\end{document}